# Most spin-1/2 transition-metal ions do have single ion anisotropy

Jia Liu[1], Hyun-Joo Koo[2], Hongjun Xiang[3,*], Reinhard K. Kremer[4] and Myung-Hwan Whangbo[1,*]

[1] Department of Chemistry, North Carolina State University, Raleigh, North Carolina 27695, USA

[2] Department of Chemistry and Research Institute for Basic Sciences, Kyung Hee University, Seoul 130-701, Republic of Korea

[3] Key Laboratory of Computational Physical Sciences (Ministry of Education), State Key Laboratory of Surface Physics, and Department of Physics, Fudan University, Shanghai 200433, People's Republic of China

[4] Max-Planck-Institut für Festkörperforschung, Heisenbergstrasse 1, D-70569 Stuttgart, Germany

**Abstract**

The cause for the preferred spin orientation in magnetic systems containing spin-1/2 transition-metal ions was explored by studying the origin of the easy-plane anisotropy of the spin-1/2 $Cu^{2+}$ ions in $CuCl_2 \cdot 2H_2O$, $LiCuVO_4$, $CuCl_2$ and $CuBr_2$ on the basis of density functional theory and magnetic dipole-dipole energy calculations as well as a perturbation theory treatment of the spin-orbit coupling. We find that the spin orientation observed for these spin-1/2 ions is not caused by their anisotropic spin exchange interactions, nor by their magnetic dipole-dipole interactions, but by the spin-orbit coupling associated with their crystal-field split d-states. Our study also predicts in-plane anisotropy for the $Cu^{2+}$ ions of $Bi_2CuO_4$ and $Li_2CuO_2$. The results of our investigations dispel the mistaken belief that magnetic systems with spin-1/2 ions have no magnetic anisotropy induced by spin-orbit coupling.

## I. Introduction

It is commonly believed that magnetic systems made up of spin-1/2 transition-metal ions have no magnetic anisotropy arising from spin-orbit coupling (SOC).[1] Thus the preferred spin orientation observed for such systems has been accounted for by invoking anisotropic spin exchange (ASE) or magnetic dipole-dipole (MDD) interactions between the spin-1/2 ions, as carried out, for example, by Moriya and Yoshida in their study of the $Cu^{2+}$ (S = 1/2) spin orientation in $CuCl_2 \cdot 2H_2O$ six decades ago.[1a] This conventional belief arises from the effective spin approximation,[2] in which magnetic ions are treated as spin-only ions and the effect of their unquenched orbital moments is included into anisotropic g-factors.[2,3] For a magnetic ion such as $Cu^{2+}$ with nondegenerate magnetic orbital, the effective spin approximation reduces the SOC Hamiltonian $\hat{H}_{SO} = \lambda\hat{S}\cdot\hat{L}$ to the zero-field Hamiltonian,

$$\hat{H}_{zf} = D\left(\hat{S}_z^2 - \tfrac{1}{3}\hat{S}^2\right) + \tfrac{1}{2}E\left(\hat{S}_+\hat{S}_+ + \hat{S}_-\hat{S}_-\right), \quad (1)$$

where the constants D and E are related to the unquenched orbital angular momenta along the three local x-, y- and z-directions of the ion, which may be denoted as $\delta L_{||x}$, $\delta L_{||y}$ and $\delta L_{||z}$, respectively, although they are not quantities to calculate or measure directly. If we take the local z-axis along the "axial" direction of a magnetic ion (located at a certain coordinate site), then the local x- and y-axes lie in the "equatorial" plane. In terms of the unquenched orbital momenta, the constant D is expressed as $D \propto \lambda^2(\delta L_{||z} - \delta L_{\perp z})$, where $\delta L_{\perp z} = (\delta L_{||x} + \delta L_{||y})/2$, and describes the difference between the "axial" and "equatorial-plane" components.[3] The constant E, written as $E \propto \lambda^2(\delta L_{||x} - \delta L_{||y})$, describes the anisotropy within the equatorial plane.[3] The three states of an S=1 ion, $|1,-1\rangle$, $|1,0\rangle$ and $|1,+1\rangle$, are split by $\hat{H}_{zf}$ hence explaining the magnetic anisotropy of such a magnetic ion in the absence of an external magnetic field. In contrast, the two states of an S=1/2 ion, $|\uparrow\rangle = |\tfrac{1}{2},+\tfrac{1}{2}\rangle$ and $|\downarrow\rangle = |\tfrac{1}{2},-\tfrac{1}{2}\rangle$, are not split by $\hat{H}_{zf}$ because $\langle\uparrow|\hat{H}_{zf}|\downarrow\rangle = 0$, as required by the Kramers degeneracy theorem for odd-spin ions.[4]

Thus, as far as the zero-field Hamiltonian is concerned, any S=1/2 ion cannot have magnetic anisotropy induced by SOC. So far, however, it has not been questioned whether this conclusion arising solely from the model Hamiltonian $\hat{H}_{zf}$ is relevant at all for describing real magnetic compounds (e.g., molecules and solids) possessing S=1/2 ions, which are experimentally found to exhibit magnetic anisotropy. That the effective spin approximation cannot describe the magnetocrystalline anisotropy of systems with S=1/2 ions has been pointed out earlier in the study [5] of $Y_2V_2O_7$ containing $V^{4+}$ ($d^1$, S = 1/2) ions at slightly-distorted octahedral sites. Surprisingly, there has been no systematic study concerning whether or not magnetic compounds with S=1/2 ions have SOC-driven magnetic anisotropy.

In this work we probe this question by examining the cause for the observed spin orientations of the $Cu^{2+}$ (S = 1/2) ions in $CuCl_2{\cdot}2H_2O$, $LiCuVO_4$, $CuCl_2$, $CuBr_2$, $Bi_2CuO_4$ and $Li_2CuO_2$, in which the $Cu^{2+}$ ions are located at $CuL_4$ (L = O, Cl, Br) square-planar sites. On the basis of density functional theory (DFT) calculations, MDD energy calculations, and a perturbation theory analysis using $\hat{H}_{SO}$ as perturbation, we show that the preferred orientation of the $Cu^{2+}$ spins in these compounds is not caused by their ASE interactions, nor by their MDD interactions. It is caused by their single-ion anisotropy (SIA), namely, their SOC-driven magnetic anisotropy, which results when the crystal-field-split d-states of each magnetic ion $Cu^{2+}$ interact under the action of SOC.

## II. Computational Details

Our spin-polarized DFT calculations employed the projector augmented wave method encoded in the Vienna ab initio simulation package (VASP) [6] and the generalized gradient approximation of Perdew, Burke and Ernzerhof [7] for the exchange-correlation functionals with the plane wave cutoff energies of 400 eV, and the threshold of self-consistent-field (SCF) energy convergence of $10^{-6}$ eV. To describe the electron correlation associated with the Cu 3d states, we employed the DFT plus on-site repulsion U (DFT+U) method of Liechtenstein et al.[8] with an effective $U_{eff} = U – J = 4$, 5 and 6 eV on the Cu atom and the exchange-correction J = 1 eV since our calculations include noncollinear magnetic structures as well.[9] Preferred spin orientations were examined by performing DFT+U calculations including SOC (DFT+U+SOC) with the threshold of SCF energy convergence of $10^{-8}$ eV, and also by MDD energy calculations.[10] In summing the MDD interactions between various pairs of spin sites, we employed the Ewald summation method.[11] In the VASP code the effect of SOC is treated by the second-variation method.[12] The DFT+U+SOC approach the present work employs has been successfully used to explain the magnetic anisotropy of, for example, $SrFeO_2$ [13] and $TbMnO_3$.[14]

## III. Results and Discussion

Before we present our results in the following, it is worthwhile to briefly mention the various terms of a spin Hamiltonian employed to describe the magnetic properties of a material with localized electrons. When there is no SOC, the spin Hamiltonian is expressed as

$$\hat{H}_{spin} = -\sum_{i<j} J_{ij}\hat{S}_i \cdot \hat{S}_j = \sum_{i<j} J_{ij}(\hat{S}_{ix}\hat{S}_{jx} + \hat{S}_{iy}\hat{S}_{jy} + \hat{S}_{iz}\hat{S}_{jz}) \qquad (2a)$$

with the isotropic Heisenberg exchange interactions $J_{ij}$ between spins at the sites i and j. In the presence of SOC, these interactions can become anisotropic, i.e., they may become different along the x-, y- and z-directions ($J_{ij-x} \neq J_{ij-y} \neq J_{ij-z}$). In such a case, Eq. (2a) should be rewritten as

$$\hat{H}_{spin} = -\sum_{i<j}(J_{ij-x}\hat{S}_{ix}\hat{S}_{jx} + J_{ij-y}\hat{S}_{iy}\hat{S}_{jy} + J_{ij-z}\hat{S}_{iz}\hat{S}_{jz}) \qquad (2b)$$

With the sign convention chosen for the spin Hamiltonians in Eqs. (2a) and (2b), the spin exchange constants (i.e., $J_{ij}$, $J_{ij-x}$, $J_{ij-y}$, $J_{ij-z}$) are antiferromagnetic (AFM) if they are negative, but ferromagnetic (FM) if they are positive.

In the presence of SOC, a spin Hamiltonian can be generally expressed as[15]

$$\hat{H}_{spin} = -\sum_{i<j}(J_{ij-x}\hat{S}_{ix}\hat{S}_{jx} + J_{ij-y}\hat{S}_{iy}\hat{S}_{jy} + J_{ij-z}\hat{S}_{iz}\hat{S}_{jz}) + \sum_{i<j}\vec{D}_{ij}\cdot(\hat{S}_i \times \hat{S}_j) + \sum_i A_i S_{iz}^2 \qquad (2c)$$

where the second term the anisotropic antisymmetric exchange (i.e., Dzyaloshinskii-Moriya) interactions between spins at the sites i and j. The third term represents the SOC-driven magnetic anisotropy of the spin at each site i, which occurs when the crystal-field-split d-states of a $Cu^{2+}$ ion at the site i interact through SOC, namely, the SIA of the $Cu^{2+}$ ion at the site i. Here the local z-axis at each $Cu^{2+}$ ion site is taken along the direction perpendicular to its equatorial plane of the axially-elongated $CuL_6$ (L = O, halogen) octahedron. It should be emphasized that the third term of Eq. (2c) employs the z-component $S_z$ of the classical spin vector $\vec{S}_i$ instead of the quantum mechanical operator $\hat{S}_{iz}$. Each SIA term $A_i S_{iz}^2$ shows that the spin $\vec{S}_i$ will orient along the local z-axis (easy-axis anisotropy) if $A_i < 0$, but perpendicular to the local z-axis (easy-plane anisotropy) if $A_i > 0$. (A more general expression for SIA is discussed in the Supplementary Material.[16] For our discussion of the easy-axis vs. easy-plane anisotropy, the simplified expression given by the third term of Eq. (2c) is sufficient.) For various practical calculations, the spin operators $\hat{S}_i$ and $\hat{S}_j$ in the first and second terms of Eq. (2c) are treated as the classical spin vectors $\vec{S}_i$ and $\vec{S}_j$, respectively.

**A. $CuCl_2·2H_2O$**

In the crystal structure of $CuCl_2·2H_2O$, isolated $CuCl_2(OH_2)_2$ complexes form skewed stacks along the b-direction (**Fig. 1a**) (in the setting of the standard space group Pmna),[17] and in each $CuCl_2(OH_2)_2$ complex the linear O-Cu-O unit is perpendicular to the linear Cl-Cu-Cl unit. The $CuCl_2(OH_2)_2$ complexes form skewed-stack chains along the b-direction (**Fig. 1b**), and there are two chains per unit cell. The spin at each $Cu^{2+}$ site is oriented along the Cu-O bond (i.e., lies in the $CuCl_2O_2$ plane), and the $Cu^{2+}$ spins have an AFM coupling within each skewed stack but a FM coupling between adjacent stacks (**Fig. 1c**).[18] We examine the cause for this easy-plane anisotropy as described below.

(1) We design an "isolated $Cu^{2+}$ ion" model (Model I) by substituting $Mg^{2+}$ ions for $Cu^{2+}$ ions such that a (2a, 3b, 2c) supercell containing one $Cu^{2+}$ ion repeats. The local coordinate axes of an isolated $CuCl_2(OH_2)_2$ complex were chosen as depicted in **Fig. 1d**, where the local x- and y-axes are taken along the Cu-Cl and Cu-O bonds, respectively. Our DFT+U+SOC calculations for the "isolated $Cu^{2+}$ ion" model with the spin orientations taken along the Cu-O bonds (i.e., ||xy) and perpendicular to the $CuCl_2O_2$ plane (i.e., ⊥xy) show that the ||xy orientation is more stable than the ⊥xy orientation (**Table I**). Because both the spin exchange and MDD interactions between adjacent $Cu^{2+}$ ions are negligible in this "isolated $Cu^{2+}$ ion" model, this result shows that the easy-plane anisotropy is an intrinsic property of an isolated $CuCl_2(OH_2)_2$ complex.

(2) We construct an "isolated $Cu^{2+}$ chain model" in which, of two chains per unit cell, all $Cu^{2+}$ ions in one chain are replaced with nonmagnetic $Mg^{2+}$ ions, and examine the ↑↓↑↓ and ↑↑↓↓ spin arrangements along the "isolated $Cu^{2+}$ chain" (to be referred to as Models II and III, respectively). For each of these two spin arrangements, we carry out DFT+U+SOC calculations for the ||xy and ⊥xy spin orientations. (In the ||xy spin orientation, the spin is aligned along the Cu-O bond, as found experimentally. See below for further discussion.) As for the spin exchange of the "isolated $Cu^{2+}$ chain", we consider only the nearest-neighbor exchange $J_{nn}$ along the chain since the next-nearest-neighbor exchange $J_{nnn}$ is negligible owing to the very long Cu…Cu distance (7.482 Å) involved. Given the AFM spin order along the stacking direction, it is reasonable to assume that the components $J_{nn-x}$, $J_{nn-y}$ and $J_{nn-z}$ of the nearest-neighbor exchange are all AFM. Given the convention that a negative spin exchange constant represents an AFM interaction, the spin exchange interaction energy per spin site for the ↑↓↑↓ arrangement is given by

$$- J_{nn-x}S_{1x}S_{2x} - J_{nn-y}S_{1y}S_{2y} - J_{nn-z}S_{1z}S_{2z}, \qquad (3)$$

where two adjacent spin sites are labeled as 1 and 2 for convenience. Then, the ||xy spin orientation is explained if $|J_{nn-x}|, |J_{nn-y}| > |J_{nn-z}|$. In the ↑↑↓↓ arrangement the sum of the nearest-neighbor spin exchange interaction energies is zero, so the ||xy and ⊥xy spin orientations would

be identical in energy if the ASE interactions were the cause for the spin orientation. However, **Table I** shows that the preference for the in-plane spin orientation is practically identical in both ↑↓↑↓ and ↑↑↓↓ arrangements. This reveals that the ASE interactions are not responsible for the easy-plane anisotropy of the $Cu^{2+}$ ion.

We use the "isolated $Cu^{2+}$ chain" model to evaluate the values of $J_{nn-x}$, $J_{nn-y}$ and $J_{nn-z}$ by using the energy-mapping analysis.[15] To obtain $J_{nn-x}$, we consider the FM and AFM spin arrangements with spins collinear to the x-direction. Representing the energies of these two state by $E(\uparrow\uparrow)_x$ and $E(\uparrow\downarrow)_x$, respectively, we obtain

$$J_{nn-x} = [E(\uparrow\uparrow)_x - E(\uparrow\downarrow)_x]/2. \quad (4a)$$

Thus $J_{nn-x}$ is determined by obtaining the energy difference, $E(\uparrow\uparrow)_x - E(\uparrow\downarrow)_x$, on the basis of DFT+U+SOC calculations. The values of $J_{nn-y}$ and $J_{nn-z}$ are similarly determined.

$$J_{nn-y} = [E(\uparrow\uparrow)_y - E(\uparrow\downarrow)_y]/2 \quad (4b)$$

$$J_{nn-z} = [E(\uparrow\uparrow)_z - E(\uparrow\downarrow)_z]/2 \quad (4c)$$

Our results summarized in **Table II** show that there is practically no anisotropy in the spin exchange constants. This also reveals that the easy-plane anisotropy of the $Cu^{2+}$ ion is not caused by the ASE interactions.

(3) We examine the MDD energies for two spin configurations, namely, the ↑↓↑↓ and ↑↑↓↓ arrangements within each skewed stack of $Cu^{2+}$ ions but with FM spin arrangement between adjacent stacks. For each of these spin arrangements, we calculate the MDD energies with the ||xy and ⊥xy spin orientations to find that the energy difference, $E_{||xy} - E_{\perp xy}$, between the two spin orientations is negligibly small (i.e., +0.005 and -0.014 meV/Cu for the ↑↓↑↓ and ↑↑↓↓ arrangements, respectively). This clearly shows that MDD interactions cannot be responsible for the easy-plane anisotropy of the $Cu^{2+}$ spin in $CuCl_2 \cdot 2H_2O$.

(4) The above discussion shows beyond any doubt that the easy-plane anisotropy of the $Cu^{2+}$ spin in $CuCl_2 \cdot 2H_2O$ is an intrinsic property of an isolated $CuCl_2(OH_2)_2$ complex that is associated with SOC. The two spin directions of interest in the ||xy spin orientation are the Cu-O and Cu-Cl bond directions. Our DFT+U+SOC calculations for the "isolated $Cu^{2+}$ chain model" with FM spin arrangement reveal that the spin orientation along the Cu-O bond is more stable than that along the Cu-Cl bond by 0.10, 0.08 and 0.07 meV/Cu for effective $U_{eff}$ = 4, 5 and 6 eV, respectively. The preference for the spin to orient along the Cu-O bond is in agreement with experiment.[18]

(5) Let us now demonstrate that the observed and calculated spin orientation of $CuCl_2(OH_2)_2$ is caused solely by SIA. By using the coordinate systems (x, y, z) and (x′, y′, z′) for the orbital and spin, respectively, the SOC Hamiltonian $\hat{H}_{SO} = \lambda \hat{S} \cdot \hat{L}$ is expressed as[3,15,19]

$$\begin{aligned}\hat{H}_{SO} = {} & \lambda \hat{S}_{z'} \left( \hat{L}_z \cos\theta + \frac{1}{2}\hat{L}_+ e^{-i\varphi} \sin\theta + \frac{1}{2}\hat{L}_- e^{i\varphi} \sin\theta \right) \\ & + \frac{\lambda}{2}\hat{S}_{+'} \left( -\hat{L}_z \sin\theta - \hat{L}_+ e^{-i\varphi} \sin^2\frac{\theta}{2} + \hat{L}_- e^{i\varphi} \cos^2\frac{\theta}{2} \right) \\ & + \frac{\lambda}{2}\hat{S}_{-'} \left( -\hat{L}_z \sin\theta + \hat{L}_+ e^{-i\varphi} \cos^2\frac{\theta}{2} - \hat{L}_- e^{i\varphi} \sin^2\frac{\theta}{2} \right)\end{aligned} \tag{5}$$

Thus $\hat{H}_{SO}$ is rewritten as $\hat{H}_{SO} = \hat{H}^0_{SO} + \hat{H}'_{SO}$, where $\hat{H}^0_{SO}$ is the spin-conserving term (i.e., the first line of Eq. 1), and $\hat{H}'_{SO}$ is the spin-non-conserving term (i.e., the second and third lines of Eq. 1). Then the preferred spin orientation (i.e., the z′ axis) is along the z-axis (i.e., θ = 0°) for the easy-axis anisotropy, but perpendicular to the z-axis (i.e., θ = 90°) for the easy-plane anisotropy. It was shown earlier that the easy-plane anisotropy of the high-spin $Fe^{2+}$ ($d^6$, S = 2) ions in $SrFeO_2$ [13] and $Sr_3Fe_2O_5$ [10] as well as the easy-axis anisotropy of the high-spin $Mn^{3+}$ ($d^4$, S = 2) ions in $TbMnO_3$ [14] and $Ag_2MnO_2$ [20] are well explained on the basis of a perturbation theory analysis by using $\hat{H}_{SO} = \lambda \hat{S} \cdot \hat{L}$ as perturbation and the crystal-field split d-states of a magnetic ion as unperturbed states.

When an occupied down-spin d-level i = $\psi_o\downarrow$ with energy $e_i$ interacts with an unoccupied down-spin d-level j = $\psi_u\downarrow$ with energy $e_j$ via the matrix element $\langle i|\hat{H}_{SO}^0|j\rangle$, the associated energy lowering is given by

$$\Delta E_{SOC} = -\frac{|\langle i|\hat{H}_{SO}^0|j\rangle|^2}{|e_i - e_j|}. \qquad (6)$$

In the following discussion we recall that, in terms of spherical harmonics $Y_\ell^m$, the angular behavior of the d-orbitals are given by $z^2 \propto Y_2^0$, $xz \propto Y_2^1 - Y_2^{-1}$, $yz \propto Y_2^1 + Y_2^{-1}$, $xy \propto Y_2^2 - Y_2^{-2}$, and $x^2\text{–}y^2 \propto Y_2^2 + Y_2^{-2}$. In determining the preferred spin orientation, the most important interaction is the one with the smallest energy gap ($e_i - e_j$). According to the split Cu 3d states of $CuCl_2{\cdot}2H_2O$ (**Fig. 1d**), the empty $(x^2\text{–}y^2)\downarrow$ has the smallest energy gap with the filled $xz\downarrow$ state. (It should be pointed out that the split Cu 3d states determined from the DFT+U calculations do not violate Kramers degeneracy theorem. It is only by convention that the up-spin states are chosen as the majority-spin states, and the down-spin states as the minority-spin states. The alternative choice in which the down-spin states are chosen as the majority-spin states is equally valid, thereby maintaining the doublet degeneracy of a spin-1/2 system.) The magnetic orbital quantum number *m* of xz differs from that of $x^2\text{–}y^2$ by 1 so that the matrix element $\langle xz\downarrow|\hat{H}_{SO}^0|(x^2-y^2)\downarrow\rangle$ can be nonzero via the $\hat{L}_+$ and $\hat{L}_-$ terms of $\hat{H}_{SO}^0$. Since these terms have the coefficient $\sin\theta$, the interaction between the $(x^2\text{–}y^2)\downarrow$ and $xz\downarrow$ states predicts easy-plane anisotropy (i.e., $\theta = 90°$). This matrix element also predicts the spin orientation along the Cu-O bond; given $\theta = 90°$, the $\langle xz\downarrow|\hat{H}_{SO}^0|(x^2-y^2)\downarrow\rangle$ term is rewritten as

$$\left\langle xz\downarrow\left|\hat{H}_{SO}^{0}\right|(x^2-y^2)\downarrow\right\rangle \propto \left\langle xz\left|\hat{L}_x\cos\phi+\hat{L}_y\sin\phi\right|x^2-y^2\right\rangle \propto \sin\phi, \qquad (7)$$

because $\left\langle xz\left|\hat{L}_x\right|x^2-y^2\right\rangle = 0$ and $\left\langle xz\left|\hat{L}_y\right|x^2-y^2\right\rangle = -i$.[3] Consequently, the matrix element $\left\langle xz\downarrow\left|\hat{H}_{SO}^{0}\right|(x^2-y^2)\downarrow\right\rangle$ is maximized for $\phi$ = 90°, i.e., the preferred spin orientation in the local xy-plane is along the Cu-O bond (**Fig. 1d**). In short, the easy-plane anisotropy of the $Cu^{2+}$ ions in $CuCl_2·2H_2O$ is solely explained by the SOC associated with its crystal-field split d-states, and is hence caused by SIA.

**B. $CuCl_2$, $CuBr_2$ and $LiCuVO_4$**

We now demonstrate that SIA is also the origin of the easy-plane anisotropy for the $Cu^{2+}$ ions in $CuCl_2$,[21,22] $CuBr_2$ [23] and $LiCuVO_4$,[24] in which the square planar $CuL_4$ units (L = Cl, Br, O) share their opposite edges to form $CuL_2$ ribbon chains (**Fig. 2a-c**). Neutron powder and single crystal diffraction studies showed that these $Cu^{2+}$ spins prefer to lie in the planes of the $CuL_4$ units (i.e., easy-plane anisotropy). Furthermore, the spins in each $CuL_2$ ribbon chain has an incommensurate spin-spiral arrangement (**Fig. 2d**) due to the spin frustration arising from the competition of the FM nearest-neighbor spin exchange $J_{nn}$ and the AFM next-nearest-neighbor spin exchange $J_{nnn}$.[15,22-24] In the spin-spiral arrangements neighboring moments make an angle of ~90°. We examine the cause for this easy-plane anisotropy as described below.

(1) To determine if this anisotropy is caused by the ASE interactions, we consider three different models of spin arrangements in each $CuL_2$ ribbon chain:

a) One is the "isolated $Cu^{2+}$ ion" model in which, of two ribbon chains per unit cell, all $Cu^{2+}$ ions of one chain are replaced with $Mg^{2+}$ ions while the other chain is converted to a $(Cu^{2+}$-$Mg^{2+}$-$Mg^{2+}$-$Mg^{2+})_\infty$ chain (**Fig. 2e**).

b) The second model has the ↑↓↑↓ collinear arrangement of $Cu^{2+}$ spins in each ribbon chain (**Fig. 2f**).

c) The third model has the ↑→↓← spin-spiral arrangement of $Cu^{2+}$ spins in each ribbon chain (**Fig. 2d**).

For each of these spin arrangements, we carry out DFT+U+SOC calculations for the ||xy and ⊥xy spin orientations to determine their relative energies. In the case of the spin-spiral arrangement (**Fig. 2d**), the ⊥xy spin orientation means that the plane of the spin spiral is perpendicular to the ribbon plane. In the ⊥xy spin spiral, therefore, each spin direction is ±45° away from the local ±z-axis. Our calculations for the "isolated $Cu^{2+}$ ion" model show that the ||xy orientation is more stable than the ⊥xy orientation (**Table III**), indicating that the easy-plane anisotropy is an intrinsic property of an isolated $CuL_4$ unit. The relative energies, $E_{||xy} - E_{\perp xy}$, obtained from the ↑↓↑↓ chain model are practically identical to those obtained from the "isolated $Cu^{2+}$ ion" model, so the spin exchanges $J_{nn}$ and $J_{nnn}$ are not responsible for the easy-plane anisotropy. It is reasonable to assume that the components $J_{nn\text{-}x}$, $J_{nn\text{-}y}$ and $J_{nn\text{-}z}$ are all FM because $J_{nn}$ is FM, while the components $J_{nnn\text{-}x}$, $J_{nnn\text{-}y}$ and $J_{nnn\text{-}z}$ are all AFM because $J_{nnn}$ is AFM. Then, the preference for the ||xy spin orientation is explained if $J_{nn-x}$, $J_{nn-y} > J_{nn-z}$ and/or if $|J_{nnn-x}|$, $|J_{nnn-y}| > |J_{nnn-z}|$. The ↑↓↑↓ spin arrangement does not fulfil these conditions, but it shows the preference for the ||xy spin orientation, just as the "isolated $Cu^{2+}$ ion" model predicts. In the case of the spin-spiral arrangement as well, the ||xy spin orientation is favored over the ⊥xy spin orientation (**Table III**). However, the energy difference, $E_{||xy} - E_{\perp xy}$, for the spin-spiral arrangement is approximately half of that found for the ↑↓↑↓ collinear spin arrangement. This is so because, for each spin of the out-of-plane spin spiral arrangement, only 50% has the ||z spin

component. In short, the ASE interactions are not responsible for the observed easy-plane anisotropy.

(2) Our MDD energy calculations using the ↑↓↑↓ spin arrangement in each $CuL_2$ ribbon chain show that the ⊥xy spin orientation is more stable than the ||xy spin orientation by 0.07, 0.07 and 0.01 meV/Cu for $LiCuVO_4$, $CuCl_2$ and $CuBr_2$, respectively. However, these MDD interactions are too weak to influence the spin orientation.

(3) We now examine the crystal-field split Cu 3d states of $CuCl_2$, $CuBr_2$ and $LiCuVO_4$ as well as their interaction under SOC. As a representative example, **Fig. 2g** shows the PDOS plots calculated for $CuCl_2$ (For the PDOS plots of $LiCuVO_4$ and $CuBr_2$, see **Fig. S1** and **S2** of the Supplementary Material[25]). We note that $\langle x^2-y^2|\hat{L}_\mu|x^2-y^2\rangle = 0$ for μ = x, y and z,[3] and $\langle z^2\downarrow|\hat{H}_{SO}|(x^2-y^2)\downarrow\rangle = 0$ because the magnetic quantum number *m* of $z^2$ differs from those of $x^2–y^2$ by 2. Consequently, **Fig. 2g** shows that the preferred spin orientation is determined by the interaction terms $\langle xy\downarrow|\hat{H}_{SO}|(x^2-y^2)\downarrow\rangle$ and $\langle (xz/yz)\downarrow|\hat{H}_{SO}|(x^2-y^2)\downarrow\rangle$. The filled $(xz/yz)\downarrow$ states are closer to the empty $(x^2-y^2)\downarrow$ states than are the filled $xy\downarrow$ states so that the ||xy spin orientation is favored over the ⊥xy spin orientation, in agreement with experiment. The in-plane anisotropy observed for $CuBr_2$ and $LiCuVO_4$ is similarly explained. Thus, just as found for $CuCl_2{\cdot}2H_2O$, the easy-plane anisotropy of the S=1/2 ions $Cu^{2+}$ in $CuCl_2$, $CuBr_2$ and $LiCuVO_4$ is caused by their SIA.

### C. $Bi_2CuO_4$ and $Li_2CuO_2$

$Bi_2CuO_4$ consists of $CuO_4$ square planes stacked to form $CuO_4$ chains along the c-direction (**Fig. 3a**). Adjacent $CuO_4$ units in each chain are staggered, and adjacent $CuO_4$ chains are interlinked by $Bi^{3+}$ ions, which form $BiO_4$ sawhorse units (not shown for simplicity). As summarized in **Table IV**, our DFT+U+SOC calculations show easy-plane anisotropy. Easy-axis anisotropy was reported for $Bi_2CuO_4$ in some experiments,[26,27] but easy-plane anisotropy was concluded from other experiments.[28] These conflicting findings may be related to the quality of the $Bi_2CuO_4$ samples employed as can be seen from their bulk magnetic properties; the Curie-Weiss temperature of $Bi_2CuO_4$ is approximately -40 K in the studies reporting easy-axis anisotropy,[26] but was found close to -100 K in the studies reporting easy-plane anisotropy.[28] Furthermore, the neutron diffraction study concluding easy-axis anisotropy employed a powder sample,[26] but that concluding easy-plane anisotropy a single crystal sample.[28a] The magnetic diffraction intensities are more sensitive to the moment direction for a single crystal than for a powder sample.

$Li_2CuO_2$ consists of $CuO_2$ ribbon chains of edge-sharing $CuO_4$ square planes (**Fig. 3b**), and the neutron diffraction studies [29] show that the $Cu^{2+}$ spins in each ribbon chain have a FM coupling, essentially with easy-axis anisotropy (the reported spin orientation is slightly tilted away from the local z-axis of each $CuO_4$ square plane, i.e., from the crystallographic c-direction, in ref. 29b). The absence of a spin spiral order and the adoption of an FM order in the $CuO_2$ ribbon chains of $Li_2CuO_2$ have been explained as a consequence of order by disorder induced by next-nearest-neighbor interchain spin exchange interactions.[30] Just as found for all other S=1/2 systems discussed above, our DFT+U+SOC calculations predict easy-plane anisotropy for $Li_2CuO_2$ (**Table V**), in apparent contradiction to the experimental finding.[29] However, it should be pointed out that the order-by-disorder state is not properly simulated by a few ordered spin states employed in our DFT+U+SOC calculations. Further studies are necessary to resolve this discrepancy.

## IV. Concluding remarks

In summary, the observed easy-plane anisotropy of the S=1/2 $Cu^{2+}$ ions in $CuCl_2{\cdot}2H_2O$, $LiCuVO_4$, $CuCl_2$ and $CuBr_2$ is not caused by their ASE interactions, nor by their MDD interactions. Our study clearly reveals that its origin is the SIA, namely, the SOC-driven magnetic anisotropy, which results when the crystal-field-split d-states of each magnetic ion $Cu^{2+}$ interact under the action of SOC. An S=1/2 ion should possess SIA unless the magnetic orbital accommodating an unpaired electron is triply degenerate as found in $Ba_2NaOsO_6$,[31] so most S=1/2 ions should exhibit SIA. It is necessary to dispel the mistaken belief that magnetic systems with spin-1/2 ions have no SOC-induced magnetic anisotropy. In superconducting precursor $La_2CuO_4$, calculations show the $Cu^{2+}$ ions to possess easy-plane anisotropy,[5] and the presence of SIA at each $Cu^{2+}$ site is not prevented by any symmetry consideration. Thus, it would be interesting to see if the SIA of $Cu^{2+}$ ions affects the spin fluctuation in cuprates [32] believed to be the key to understanding their superconductivity.

## Acknowledgments

This research was supported by the computing resources of the NERSC center and the HPC center of NCSU. Work at Fudan was partially supported by NSFC, FANEDD, NCET-10-0351, Research Program of Shanghai Municipality and MOE, the Special Funds for Major State Basic Research, Program for Professor of Special Appointment (Eastern Scholar), and Fok Ying Tung Education Foundation.

e-mails: hxiang@fudan.edu.cn, mike_whangbo@ncsu.edu

Table I. The energy (in meV/Cu) of the ||xy spin orientation with respect to that of the ⊥xy spin orientation, $E_{||xy} - E_{\perp xy}$, obtained from the DFT+U+SOC calculations for three models of $CuCl_2 \cdot 2H_2O$.[a]

| $U_{eff}$ | 4 eV | 5 eV | 6 eV |
|---|---|---|---|
| Model I[b] | -0.36 | -0.32 | -0.27 |
| Model II[c] | -0.23 | -0.20 | -0.17 |
| Model III[d] | -0.23 | -0.20 | -0.18 |

[a] In the ||xy spin orientation, the spin is parallel to the Cu-O bond.

[b] "Isolated $Cu^{2+}$ ion" model

[c] "Isolated $Cu^{2+}$ chain" model with ↑↓↑↓ spin arrangement

[d] "Isolated $Cu^{2+}$ chain" model with ↑↑↓↓ spin arrangement

Table II. The values of the three components $J_{nn-x}$, $J_{nn-y}$ and $J_{nn-z}$ (in meV) of the nearest-neighbor spin exchange $J_{nn}$ in $CuCl_2 \cdot 2H_2O$ determined from DFT+U+SOC calculations.

| | $U_{eff}$ = 4 eV | $U_{eff}$ = 5 eV | $U_{eff}$ = 6 eV |
|---|---|---|---|
| $J_{nn-x}$ | -1.335 | -1.090 | -0.885 |
| $J_{nn-y}$ | -1.335 | -1.090 | -0.890 |
| $J_{nn-z}$ | -1.335 | -1.095 | -0.890 |

Table III. The relative energies, $E_{\parallel xy} - E_{\perp xy}$, in meV/Cu obtained from DFT+U+SOC calculations for the "isolated $Cu^{2+}$ ion" model, the ↑↓↑↓ chain model (in parenthesis), and the ↑→↓← spin-spiral chain model (in square bracket).

| | $U_{eff}$ = 4 eV | $U_{eff}$ = 5 eV | $U_{eff}$ = 6 eV |
|---|---|---|---|
| $LiCuVO_4$ | -0.35[a], (-0.36)[b], [-0.18][c] | -0.31[a], (-0.31)[b], [-0.17][c] | -0.26[a], (-0.27)[b], [-0.14][c] |
| $CuCl_2$ | -0.29, (-0.26), [-0.15] | -0.23, (-0.22), [-0.13] | -0.21, (-0.19), [-0.12] |
| $CuBr_2$ | -0.13, (-0.16), [-0.07] | -0.11, (-0.14), [-0.06] | -0.10, (-0.12), [-0.05] |

[a] The "isolated $Cu^{2+}$ ion" model

[b] The ↑↓↑↓ chain model

[c] The ↑→↓← spin-spiral chain model

Table IV. The relative energies in meV/Cu obtained for the three spin orientations of $Bi_2CuO_4$ from DFT+U+SOC calculations for the experimentally observed AFM state.[a]

| Spin orientation | $U_{eff}$ = 4 eV | $U_{eff}$ = 6 eV |
|---|---|---|
| //a | -0.56 | -0.38 |
| //b | -0.56 | -0.38 |
| //c | 0 | 0 |

[a] The $Cu^{2+}$ spins in each $CuO_4$ chain are antiferromagnetically coupled, and so are those between adjacent $CuO_4$ chains such that, in **Fig. 3**, the $Cu^{2+}$ ion of the upper $CuO_4$ unit in the left $CuO_4$ chain makes an AFM coupling with the corresponding one in the right $CuO_4$ chain.

Table V. The relative energies, $E_{\|xy} - E_{\perp xy}$, in meV/Cu of $Li_2CuO_2$ obtained from DFT+U+SOC calculations for the "isolated $Cu^{2+}$ ion", the ↑↓↑↓ chain, and the ↑↑↓↓ chain models.

| | $U_{eff}$ = 4 eV | $U_{eff}$ = 6 eV |
|---|---|---|
| "isolated $Cu^{2+}$ ion" | 0.34 | 0.28 |
| ↑↓↑↓ chain | 0.25 | 0.23 |
| ↑↑↓↓ chain | 0.30 | 0.24 |

**Figure captions**

Figure 1. (Color online) (a) A perspective view of the crystal structure of $CuCl_2 \cdot 2H_2O$ (blue circle: Cu, yellow circle: Cl, red circle: O, white circle: H). (b) A schematic view of an isolated $CuCl_2(OH_2)_2$ chain along the b-direction. (c) The magnetic structure of $CuCl_2 \cdot 2H_2O$, where the filled and unfilled circles represent up-spin and down-spin $Cu^{2+}$ sites, respectively. (d) The local coordinate system chosen for an isolated $CuCl_2(OH_2)_2$ complex with the crystal-field split Cu 3d states. (e) The PDOS plots calculated for the Cu 3d states of $CuCl_2 \cdot 2H_2O$ using the "isolated $Cu^{2+}$ ion" model, where the horizontal axis refers to energy in eV (with the Fermi level $E_F = 0$), and the vertical axis the PDOS in states/eV/Cu (with the up-spin and down-spin states represented by positive and negative numbers, respectively).

Figure 2. (Color online) (a) Arrangement of $CuL_2$ ribbon chains (along the crystallographic b-direction) in $CuL_2$ (L = Cl, Br), where the Cu and L atoms are represented by blue and green spheres, respectively. (b) Arrangement of $CuO_2$ ribbon chains (along the crystallographic b-direction) in $LiCuVO_4$, where the Cu and O atoms are represented by blue and red spheres, respectively. (c) A schematic view of a $CuL_2$ (L = Cl, Br, O) ribbon chain found in $CuCl_2$, $CuBr_2$ and $LiCuVO_4$. (d) A schematic view of the ↑→↓← spin-spiral arrangement in the $CuL_2$ ribbon chain, for the case when the spin-spiral plane is the xy-plane. (e) An "isolated $Cu^{2+}$ ion" model, in which a chain of $Cu^{2+}$ ions is replaced with a $(Cu^{2+}\text{-}Mg^{2+}\text{-}Mg^{2+}\text{-}Mg^{2+})_\infty$ chain. (f) A schematic view of the ↑↓↑↓ collinear arrangement of $Cu^{2+}$ spins in the $CuL_2$ ribbon chain, for the case when the spins lie in the xy-plane. (g) The PDOS plots calculated for the Cu 3d states of $CuCl_2$, where the horizontal axis refers to energy in eV (with the Fermi level $E_F = 0$), and the vertical

axis the PDOS in states/eV/Cu (with the up-spin and down-spin states represented by positive and negative numbers, respectively).

Figure 3. (Color online) (a) Perspective view of the stacked $CuO_4$ square planes (along the crystallographic c-direction) in $Bi_2CuO_4$. (b) Arrangement of the $CuO_2$ ribbon chains (along the crystallographic b-direction) in $LiCuVO_4$. Here the Cu and O atoms are represented by blue and red spheres, respectively.

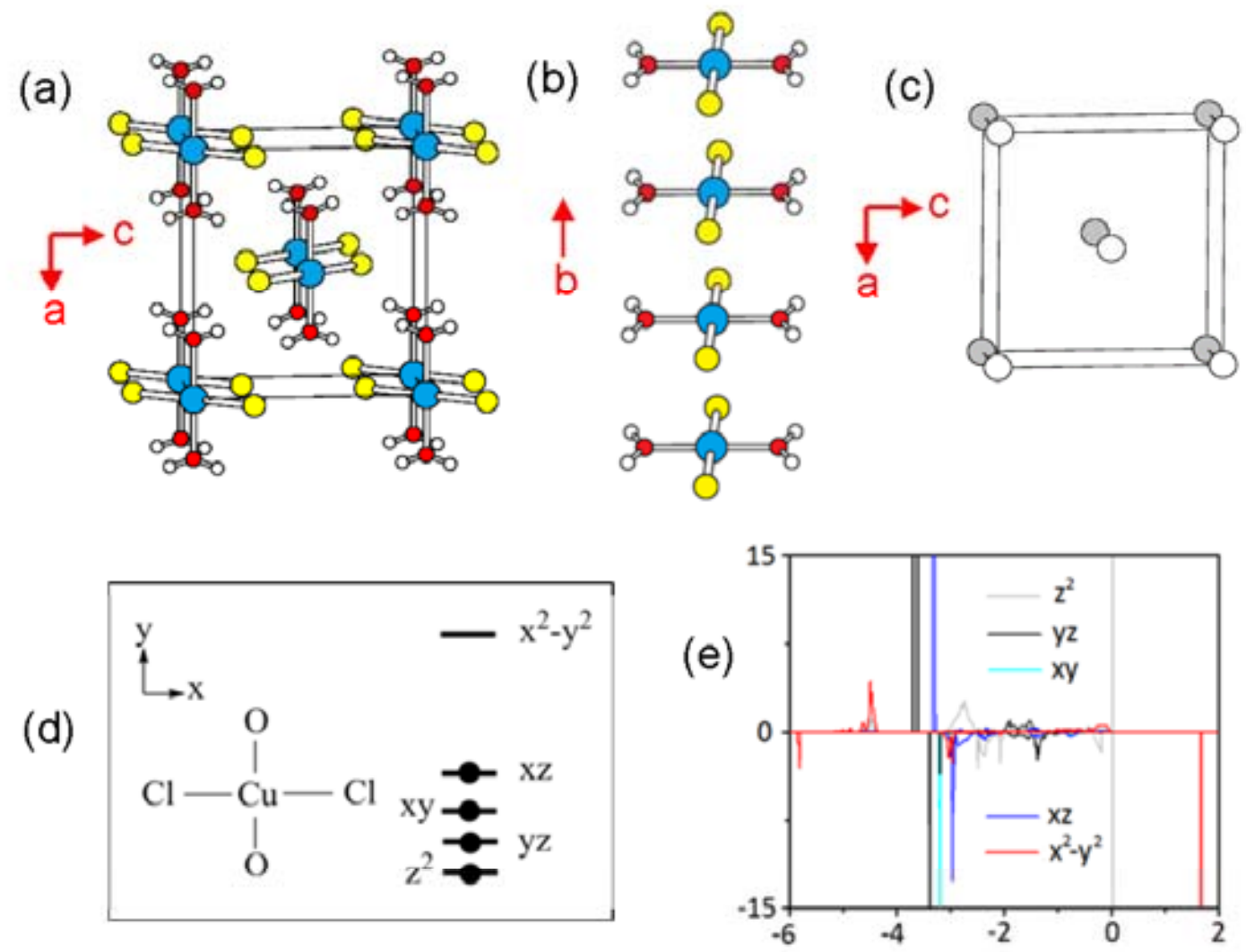
(a)
(b)
(c)
c
a
b
c
a
(d)
y
x
x2-y2
O
Cl
Cu
Cl
O
xz
xy
yz
z2
(e)
15
0
-15
z2
yz
xy
xz
x2-y2
-6
-4
-2
0
2

Fig. 1

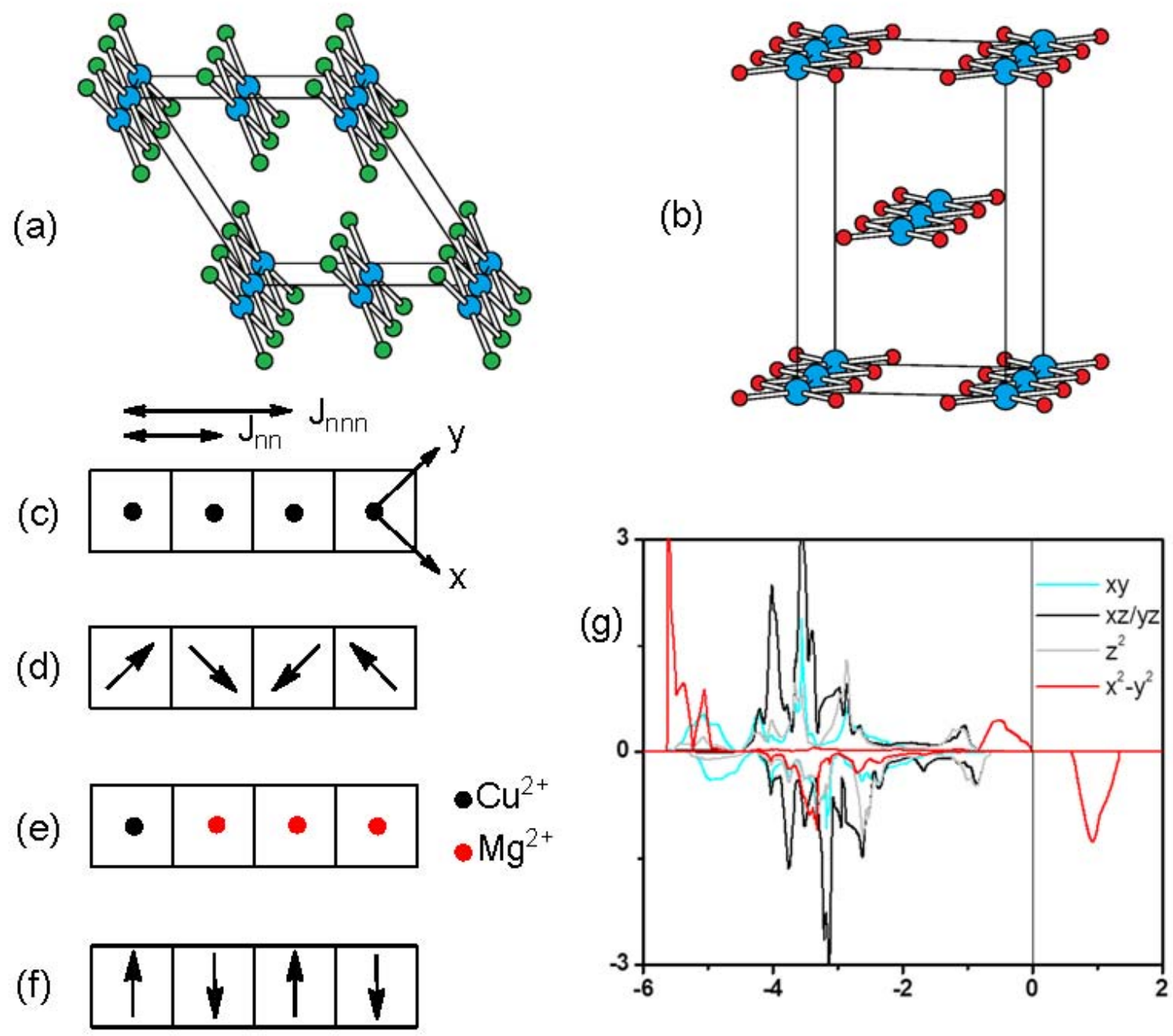

Fig. 2

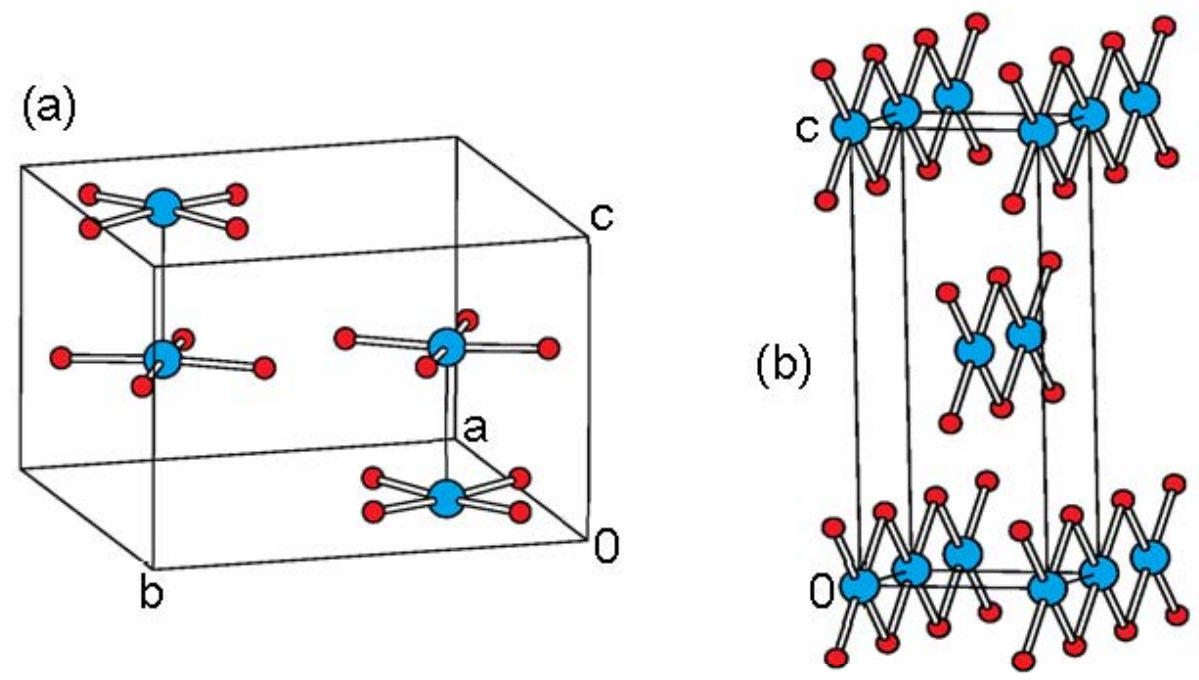

Fig. 3

Supplementary Material

**Most spin-1/2 transition-metal ions do have single ion anisotropy**

Jia Liu[1], Hyun-Joo Koo[2], Hongjun Xiang[3,*], Reinhard K. Kremer[4] and Myung-Hwan Whangbo[1,*]

[1] Department of Chemistry, North Carolina State University, Raleigh, North Carolina 27695, USA

[2] Department of Chemistry and Research Institute for Basic Sciences, Kyung Hee University, Seoul 130-701, Republic of Korea

[3] Key Laboratory of Computational Physical Sciences (Ministry of Education), State Key Laboratory of Surface Physics, and Department of Physics, Fudan University, Shanghai 200433, People's Republic of China

[4] Max-Planck-Institut für Festkörperforschung, Heisenbergstrasse 1, D-70569 Stuttgart, Germany

**1. Single ion anisotropy**

The SIA of a magnetic ion i is more generally discussed in terms of the term[1,2]

$$\mathbf{S_i} \cdot \mathbf{D} \cdot \mathbf{S_i}\,, \qquad \text{(S1)}$$

where $\mathbf{D}$ represents the second-order anisotropy tensor while $\mathbf{S_i}$ is treated as the vector operator. In the local coordinate system that diagonalizes $\mathbf{D}$ to give the diagonal elements $D_{xx}$, $D_{yy}$ and $D_{zz}$, Eq. (S1) can be replaced with[2a,3]

$$AS_{iz}^2 + B(S_{ix}^2 - S_{iy}^2)\,, \qquad \text{(S2)}$$

where $A = D_{zz} - (D_{xx} + D_{yy})/2$ and $B = (D_{xx} - D_{yy})/2$. In general, for $Cu^{2+}$ ($d^9$, S=1/2) ions that are typically found in axially-elongated octahedral coordinate sites, square pyramidal or square-planar sites, the z-axis is taken perpendicular to the $CuL_4$ square planes made up of the four equatorial ligands L. In discussing whether the preferred spin orientation is along the z-axis (i.e., the easy-axis anisotropy) or in the xy-plane (i.e., the easy-plane anisotropy), the second term of Eq. (S2) is negligible compared with the first term, and the sign of A of the first term determines the anisotropy (namely, $A < 0$ for the easy-axis anisotropy, and $A > 0$ for the easy-plane anisotropy). This justifies our use of the simplified SIA expression in Eq. (2c).

## 2. Supplementary figures

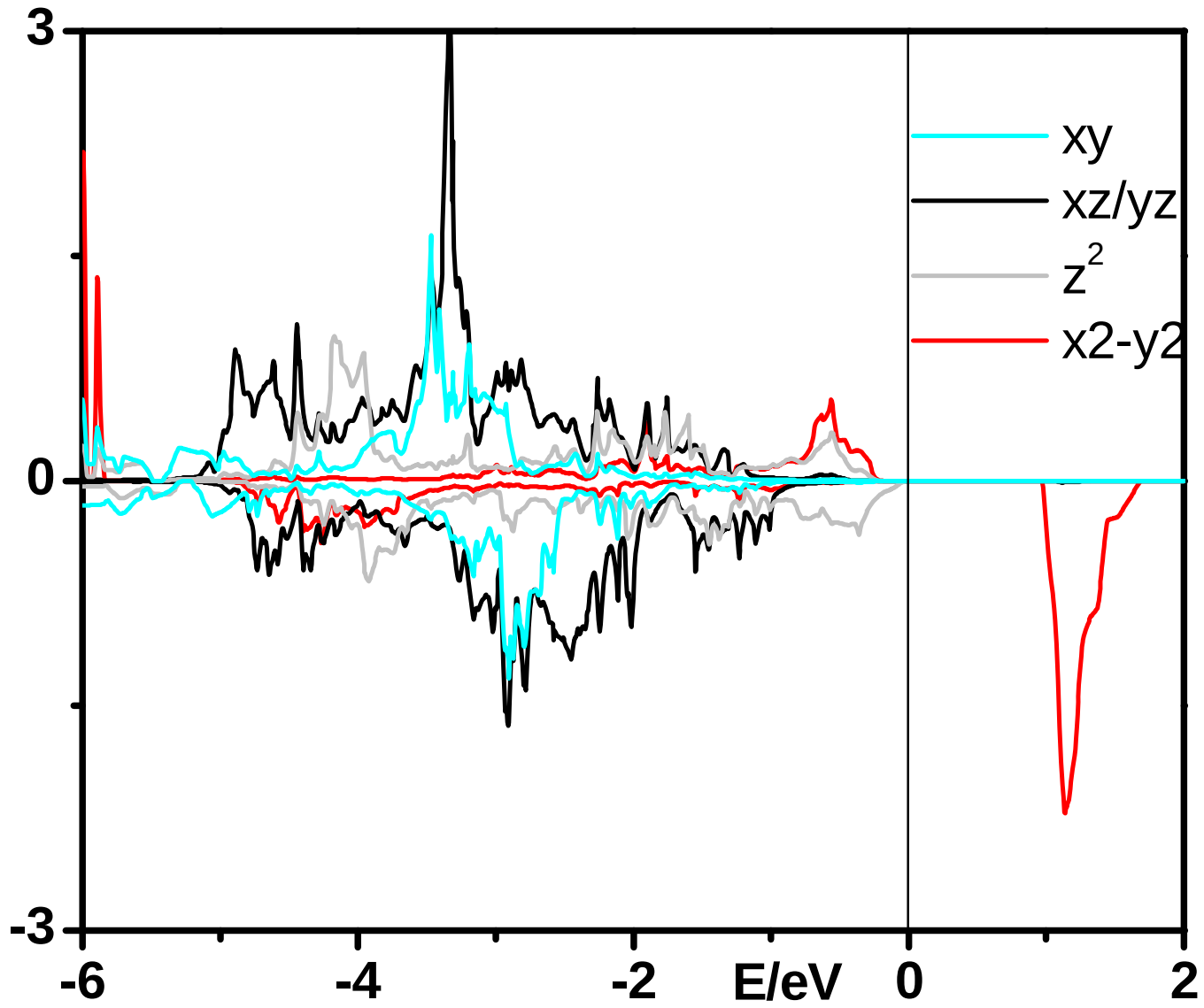

Figure S1. PDOS plots calculated for $LiCuVO_4$.

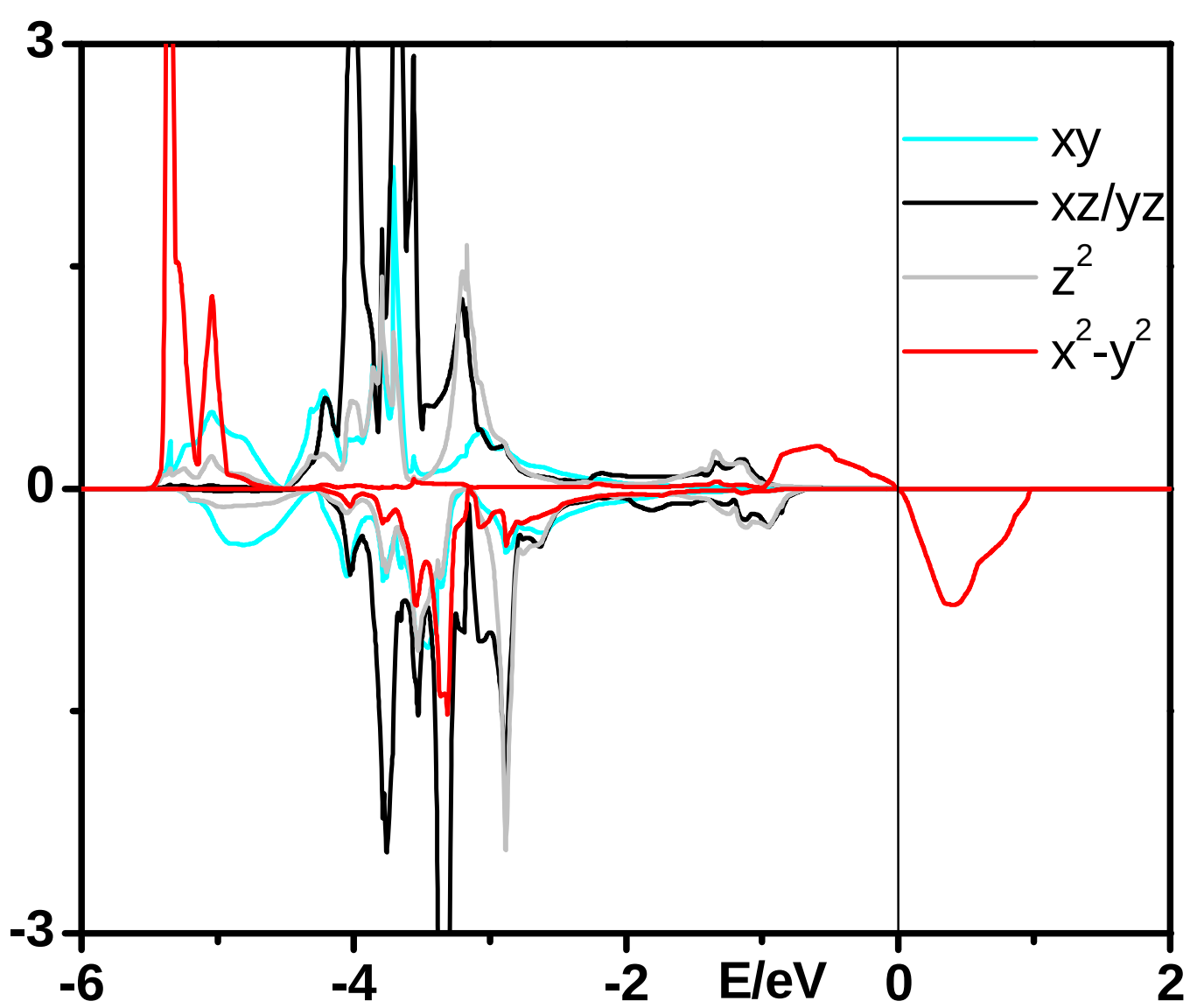

Figure S2. PDOS plots calculated for $CuBr_2$.